\documentclass[nofootinbib,superscriptaddress,twocolumn,showpacs,preprintnumbers,amsmath,amssymb]{revtex4-1}
\usepackage{hyperref}
\usepackage{graphicx}
\usepackage{grffile}
\usepackage{slashed}
\usepackage{bbm}
\usepackage{footmisc}
\usepackage{multirow}
\usepackage{amssymb}
\usepackage{amsmath}
\usepackage{color}
\usepackage{xspace}
\usepackage{caption}
\usepackage{subcaption}
\usepackage{natbib}

\usepackage{datetime}
\usepackage{color,fancybox}

\newcommand{\GeV}{\ensuremath{\,\mathrm{GeV}}\xspace}

\newcommand{\ii}{\mathord{\mathrm{i}}}

   \def\beq{\begin{equation}} \def\eeq{\end{equation}}
\def\bea{\begin{eqnarray}} \def\eea{\end{eqnarray}}

\begin{document}

\title{Higgs boson $\boldsymbol{\mathcal{CP}}$-properties of the
  gluonic contributions in Higgs plus three jet production via
  gluon fusion at the LHC}
\preprint{FTUV-14-1202\;\;IFIC-14-11\;\;KA-TP-04-2014\;\;
  LPN14-043\;\; SFB/CPP-14-11 \;\;TTK-14-07}

\author{Francisco Campanario} 
\email{francisco.campanario@ific.uv.es} 
\affiliation{Theory Division,
  IFIC, University of Valencia-CSIC, E-46100 Paterna, Valencia,
  Spain}
\affiliation{Institute for Theoretical
  Physics, KIT, 76128 Karlsruhe, Germany.}

\author{Michael Kubocz}
\email{kubocz@physik.rwth-aachen.de} \affiliation{Institut f\"ur
  Theoretische Teilchenphysik und Kosmologie,\\ RWTH Aachen
  University, D52056 Aachen, Germany}
\affiliation{Institute for Theoretical Physics, KIT, 76128 Karlsruhe, Germany.}

\begin{abstract} 
 \noindent In high energy hadronic collisions, a general
 $\mathcal{CP}$-violating Higgs boson $\Phi$ with accompanying jets
 can be efficiently produced via gluon fusion, which is mediated by
 heavy quark loops. In this letter we study the dominant sub-channel $
 gg \to \Phi ggg$ of the gluon fusion production process with triple
 real emission corrections at order $\alpha_s^5$. We go beyond the
 heavy top limit approximation and include the full mass dependence of
 the top- and bottom-quark contributions.  Furthermore, we show within
 a toy-model scenario that bottom-quark loop contributions in
 combination with large values of $\tan \beta$ can modify visibly the
 differential distributions sensitive to $\mathcal{CP}$-measurements
 of the Higgs boson particle.
\end{abstract}

\pacs{12.38.Bx, 13.85.-t, 14.65.Fy, 14.65.Ha, 14.70.Dj, 14.80.Bn} 
\keywords{Higgs boson, Standard Model, Hadronic Colliders}

\maketitle
\section{Introduction}
\label{intr}
\noindent 
Since the discovery of a new bosonic resonance with a mass in the
range of 125-126 GeV at the Large Hadron Collider~(LHC), the
measurement of its properties to validate the Standard Model~(SM)
Higgs boson hypothesis has become one of the main goals of the
scientific community~\cite{Plehn:2001nj, Djouadi:2005gi,Djouadi:2005gj,
  Cox:2010ug,Coleppa:2012eh, Freitas:2012kw,
  Englert:2012ct,Harlander:2013oja, Chang:2013cia, Djouadi:2013qya}.
Recent measurements by the ATLAS and CMS collaborations favor a spin-0
Higgs boson with a positive parity
~\cite{Aad:2012tfa,Chatrchyan:1471016}, a pure $\mathcal{CP}$-odd
scalar Higgs particle was already discarded in previous
studies~\cite{Freitas:2012kw} with more than three standard
deviations.
%
However, a scalar $\mathcal{CP}$-violating Higgs boson consisting of a
mixed state of $\mathcal{CP}$-odd and $\mathcal{CP}$-even couplings to
fermions has not still been ruled out. It can be described by the
following Lagrangian:
\begin{equation}
{\cal L}_{\rm Yukawa}=\overline{q} \, (y_q  + \ii \gamma_5
\tilde{y}_q)\, q\,  \Phi \, ,
\label{eq:ggalpha}
\end{equation}
where  $\Phi$ denotes a scalar
Higgs particle with unconstraint $\mathcal{CP}$ properties via the
assignment
\begin{equation}
\Phi= H \cos \alpha +A \sin \alpha \,,
\end{equation}
%
%
with $H$ and $A$ representing a $\mathcal{CP}$-even and
$\mathcal{CP}$-odd Higgs bosons, respectively, and $\alpha$ the
corresponding mixing angle.
Higgs production in association with two jets via gluon fusion is a
promising channel in order to measure the $\mathcal{CP}$-properties of
the Higgs particle as well as its coupling to
fermions~\cite{DelDuca:2001eu,Odagiri:2002nd}.
In general, the production of $\Phi + 2$ jets events leads to a
distinctively altered distribution of the azimuthal angle difference
between the two jets. The maximum of the distribution is found at
$\Delta \phi_{jj} = -\alpha$ and $\Delta \phi_{jj} = -\alpha \pm \pi$,
in contrast to a pure $\mathcal{CP}$-even Higgs $\Phi = H$ or
$\mathcal{CP}$-odd Higgs $\Phi = A$~\cite{Plehn:2001nj,
  Hankele:2006ma, Klamke:2007cu,Hagiwara:2009wt, Campanario:2010mi},
with maximums situated at $\Delta \phi_{jj} = 0~(\pm \pi)$ and $\Delta
\phi_{jj} = \pm \pi/2$, respectively. Thus, the azimuthal angle
distribution of $\Phi jj$ events production provides valuable
information about the mixing nature of the scalar particle.

A relevant aspect of interest is the modification of the azimuthal
angle correlation by emission of additional jets, that is, at least by
a third jet. Several analyses~\cite{DelDuca:2006hk, DelDuca:2008zz,
  Andersen:2010zx} demonstrated that the $\phi_{jj}$-correlation
survives with minimal modifications after the separation of hard
radiation from showering effects with subsequent
hadronization. Similar conclusions were obtained by a parton level
calculation with NLO
corrections~\cite{Campbell:2006xx,vanDeurzen:2013rv} to the Higgs plus
two jets process in the framework of an effective Lagrangian. In this
letter, we analyze whether the presence of additional soft radiation
may destroy the characteristic pattern observed in
Ref.~\cite{Campanario:2010mi} for $\Phi jj$ production.

For a Higgs mass lower than the top-quark mass, the total cross
section can be determined with a good accuracy via the effective
Lagrangian derived from the heavy top limit approximation
\begin{equation}
{\cal L}_{\rm eff} =
\frac{y_t}{y_t^{SM}}\cdot\frac{\alpha_s}{12\pi v} \cdot H \,
G_{\mu\nu}^a\,G^{a\,\mu\nu} +
\frac{\tilde y_t}{y_t^{SM}}\cdot\frac{\alpha_s}{8\pi v} \cdot A \,
G^{a}_{\mu\nu}\,\tilde{G}^{a\, \mu\nu}\;,
\label{eq:ggS}
\end{equation}
where $G^{a}_{\mu\nu}$ represents the gluon field strength and
$\tilde{G}^{a\, \mu\nu} = 1/2\,
G^{a}_{\rho\sigma}\,\varepsilon^{\mu\nu\rho\sigma}$ its dual. 
The validity range of the effective approach has been studied 
at LO for $\Phi jj$ production in Ref.~\cite{Campanario:2010mi}, and recently for $H
jjj$ production in Ref.~\cite{Campanario:2013mga}, for which there are
additional NLO results computed within the effective
theory~\cite{Cullen:2013saa}.

For values of the Higgs transverse momentum larger than twice the top
mass, Higgs masses bigger than the top mass and finally in models
(e.g. 2HDM, MSSM, etc.) with strong enhancement of bottom-loop
contributions by a large ratio of the two vacuum expectation values,
$v_u/v_d=\tan \beta$,
the effective Lagrangian approximation, Eq.~\ref{eq:ggS}, breaks down
and leads to unreliable predictions. Here it is necessary to switch to
the full theory with full quark mass dependence in the contributing
loops.

In this letter, we provide results for $\Phi jjj$ going beyond the
heavy top approximation, including the full mass dependence of the
top- and bottom-quark contributions at LO for the sub-process $gg \to
\Phi ggg$ which is the dominant channel, and hence, an essential piece
to compute the real emission contributions for Higgs plus two jets
production at NLO via GF within the full theory. This production
channel involves the manipulation of massive rank-5 hexagon Feynman
diagrams, which are the most complicated topologies appearing in Higgs
production in association with three jets via GF, and thus it provides
a testing ground to check the numerical stability of the full process.
This is particularly important for the numerically challenging
bottom-loop corrections, which turn out to provide the main
contributions at large values of $\tan \beta$, and hence, dominate
over the top-loop contributions.
%
%
%
Results for the full process and a detailed description of
de-correlation effects will be given in a forthcoming publication.
%
%

This letter is organized as follows. The technical details of our
implementation are presented in Section~\ref{sec:calc}. Numerical
results are shown in Section~\ref{sec:results} and finally
conclusions in Section~\ref{sec:summ}.
\begin{figure}[!ht]
\begin{center}
\includegraphics[width=1\columnwidth]{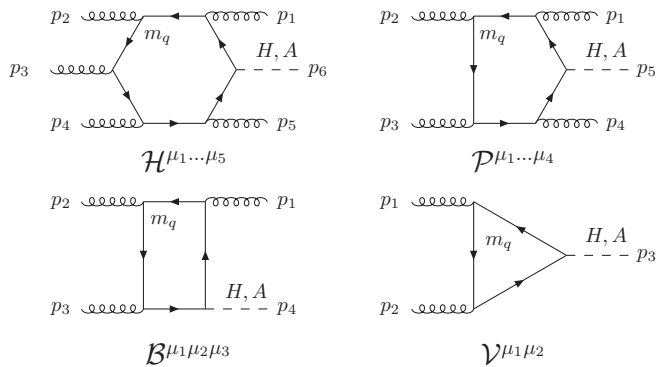}\quad
\vspace*{-1.5em}
\end{center}
\caption{Master Feynman diagrams}
\label{fig:diag}
\end{figure}
%
\section{Calculational Details}
\label{sec:calc}
The relevant subprocesses contributing to $\Phi jjj$ production are,
\begin{align}  \label{3jsubproc}
q\, q \rightarrow q\, q\,g\, \Phi , \quad q\, Q \rightarrow q\, Q\,g\,
\Phi , \nonumber \\ q\, g \rightarrow q\, g\,g\, \Phi , \quad g\ g
\rightarrow g\, g\, g\,\Phi \, .
\end{align}
In this letter, we restrict our study to the last sub-process.  In a
2HDM, Yukawa couplings to up- and down-type quarks are generally
functions of the ratio of two vacuum expectation values, $\tan \beta =
v_u/v_d$. In the scenario of the 2HDM model of type II, the Yukawa
coupling to up-type quarks is suppressed by $\cot \beta$ in contrary
to the enhancement by $\tan \beta$ of the Yukawa couplings to
down-type quarks,
\begin{align}
 \label{c:tb}
\tilde{y}^{\text{II}}_{u}= -\frac{\cot \beta}{v}m_u  
\qquad \text{and}
\qquad
\tilde{y}^{\text{II}}_{d}= -\frac{\tan \beta}{v}m_d  \; .
\end{align}
Due to this enhancement, loops with bottom-quarks can also provide
significant contributions to the total cross section as well as to
differential distributions of important observables. Thus, we take
bottom-loop corrections into account to study their phenomenological
effects and numerical behavior. In this connection we closely follow
the setup described in Ref.~\cite{Campanario:2013mga} for $Hjjj$
production. The here analyzed Higgs production process is 
available in the GGFLO MC program, which is also a part of the VBFNLO
framework~\cite{Arnold:2008rz, *Arnold:2011wj,*Arnold:2012xn}. As
customary in VBFNLO calculations, we use the effective current
approach~\cite{Hagiwara:1985yu, Hagiwara:1988pp} to evaluate loop
amplitudes. Eight master Feynman diagrams involving four $\mathcal{CP}$-even
and four $\mathcal{CP}$-odd Higgs couplings to fermions are needed. For
this letter, the four $\mathcal{CP}$-odd Higgs master integrals
depicted in Fig.~\ref{fig:diag} have been computed with the in-house
framework described in Ref.~\cite{Campanario:2011cs}-- the attached
gluons are considered to be off-shell vector currents, which allow the
attachment of further participating gluons. The
numerical evaluation of the tensor integrals follows the
Passarino-Veltman approach of Ref.~\cite{Passarino:1978jh} up to
boxes, and Ref.~\cite{Denner:2005nn}, with the scheme laid out in
Ref.~\cite{Campanario:2011cs}, for pentagons and hexagons. Scalar
integrals are computed following
Refs.~\cite{'tHooft:1978xw,Denner:1991qq}. Furthermore, the number of
diagrams to be evaluated are reduced by a factor two applying Furry's
theorem. The color factors are the same as for $Hjjj$ production and
were computed by hand and cross checked with the
program~\textsc{MadGraph}~\cite{Alwall:2007st,Alwall:2011uj}. %

To guarantee the correctness of the results, we compare first 
the amplitudes of the $\mathcal{CP}$-odd and $\mathcal{CP}$-even production
modes obtained by~\textsc{MadGraph} in the top limit approximation against
a self-made implementation of the $\Phi jjj$ production channel, implemented
also in VBFNLO. The agreement for both the $\mathcal{CP}$-odd and $\mathcal{CP}$-even production
modes is satisfied at the machine
precision level at the amplitude level and at the per mille level when
compared at the integrated cross section level against~\textsc{MadGraph}. 
Then, we compare the full and effective theory results at the integrated cross
section level for  $m_{t}=5\cdot 10^4$ GeV. The agreement is better than one
per ten thousand level. 
\begin{center}
\begin{figure*}[!thb]
\includegraphics[width=0.957\columnwidth]{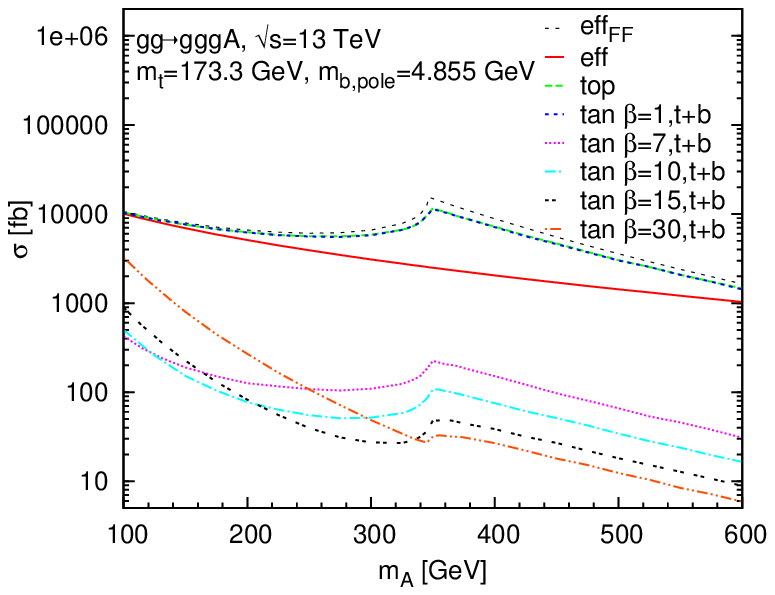}
\hfill
\includegraphics[width=0.957\columnwidth]{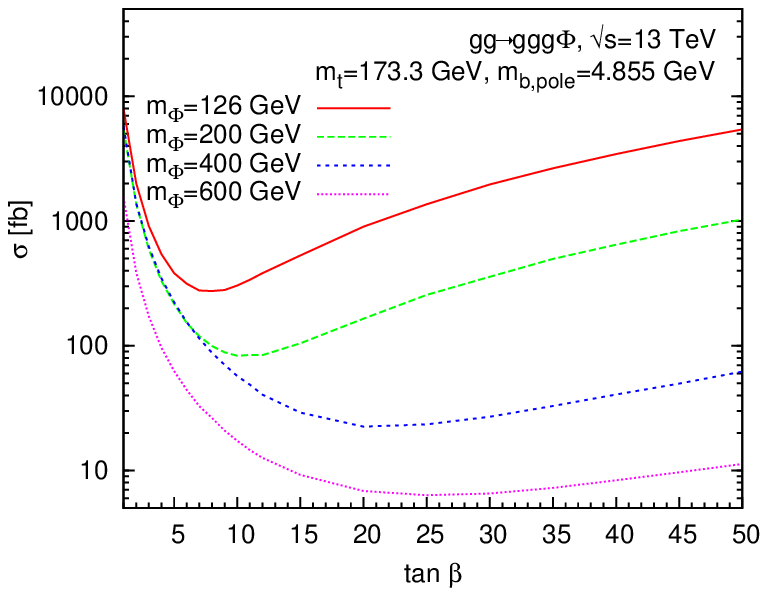}
\caption{Left: A + 3 jet cross section as a function of the
  pseudo-scalar Higgs boson mass, $m_A$, for different values of $\tan
  \beta$. Right: $\Phi $+ 3 jet cross section as a function of $\tan
  \beta$ for several values of the $\Phi$ mass. The inclusive cuts
  (IC) of Eq.~(\ref{ICuts}) are applied}
\label{fig:massscanA}
\end{figure*}
\end{center}

To control the numerical instabilities inherent to a multi-leg
calculation, we follow the procedure described in
Ref.~\cite{Campanario:2013mga}. We provide a summary here for the
sake of being self contained.
We use the Ward identities technique developed in
Ref.~\cite{Campanario:2011cs} and applied successfully in other
complex GF
processes~\cite{Campanario:2012bh,Campanario:2013mga}. These
identities allow to relate $N$-point to $N-1$-point tensor integrals
by replacing an effective current by the corresponding momentum flow. This
property is transferred to the Master integrals, hence it 
provides a strong check on the correctness of the
Master integrals. For example, a hexagon topology of tensor rank five
can be written with the help of the Ward identity as a difference of
two pentagons topologies of tensor rank four
\begin{equation}
{\cal H}^{\mu_1 \ldots \mu_5} p_{i,\mu_i} =
{\cal P}_1^{\mu_1 \ldots \hat{\mu}_i \ldots \mu_5}
- {\cal P}_2^{\mu_1 \ldots \hat{\mu}_i \ldots \mu_5},
\hspace{0.05cm} i  =1\ldots 5 \,,
\label{ward}
\end{equation}
where $\hat{\mu}_i$ denotes the corresponding vertex replaced by its
momentum $p_i$.
We construct all possible Ward identities for each physical
permutation and diagram, e.g. all five different ones for the hexagon
${\cal H}^{\mu_1\ldots \mu_5}$. 

For each phase space point and diagram, these Ward identities are
evaluated with a small additional computing effort using a cache
system.  We request a global accuracy of $\epsilon=5 \times 10^{-4}$
and reevaluate the diagram using quadruple precision if the Ward Identities are not satisfied
with the demanded accuracy.  Finally, the amplitude is set to zero and
the phase-space point discarded if the Ward identities are not
satisfied after this step.  The amount of phase-space points, which
does not pass the Ward identities after this step is statistically negligible and well
below the per mille level.

With this method, we obtain statistical error of $1\%$ in $3$ hours (top
contributions only) for the LO inclusive cross section using 
a single core of an Intel $i7$-$3970X$ processor with the Intel-ifort
compiler (version $12.1.0$). The distributions shown below are based
on multiprocessor runs with a total statistical error of up to
$0.02\%$. These precision and computing-time results set a benchmark
for comparisons with automated multi-leg calculation programs.
\section{Numerical Results}
\label{sec:results}

In this section, we present results for integrated cross sections and
selected differential distributions of important observables for the
sub-process $gg \to ggg \Phi$ at the LHC at 13 TeV center of mass
(c.m.) energy. We use the CTEQ6L1 parton distribution functions (PDFs)
\cite{Pumplin:2002vw} with the default strong coupling value
$\alpha_s(M_Z)=0.130$ and the $k_T$-jet algorithm. To avoid soft and
collinear QCD singularities, we introduce a minimal set of cuts:
\begin{align} 
\label{ICuts}
p_{T}^{j_i} > 20 \ \text{GeV}\;, \qquad |y_j| < 4.5\;, \qquad R_{jj} >
0.6 \; ,
\end{align}
where $R_{jj}$ describes the separation of the two partons in the
rapidity versus azimuthal-angle plane,
\begin{align}
R_{jj} = \sqrt{\Delta y_{jj}^2 + \phi_{jj}^2} \;,
\end{align}
with $\Delta y_{jj} = |y_{j1}-y_{j2}|$ and $\phi_{jj} =
\phi_{j1}-\phi_{j2}$. These cuts anticipate LHC detector capabilities
and jet finding algorithms and will be called ``inclusive cuts'' (IC)
in the following.
All quarks, except the bottom- and the top-quark are considered
massless. The top-quark mass is fixed at $m_t=173.3$\,GeV and
$\overline{\text{MS}}$ bottom-quark mass at
$\overline{m}_b(m_b)=4.2$\,GeV. In our setup Yukawa couplings contain
a 33-42$\%$ smaller $m_b$ than the pole mass of 4.855 GeV utilized in
the loop propagators within the Higgs-mass range of 100-600 GeV. 
Although we present a LO calculation, we have
taken into account the evolution of $m_b$ up to a reference scale (in
this case $m_H$) due to the dominance of the bottom loop contributions
at large values of $\tan \beta$. We achieved this by utilizing the
relation between the pole mass and the $\overline{\text{MS}}$ mass,
following Refs.~\cite{Spira:1997dg,Vermaseren:1997fq} within a
5-flavor scheme. The Higgs boson is produced on-shell and without
finite width effects.  Additionally, we choose $ M_Z = 91.188
\text{GeV}$, $M_W=80.386 \text{GeV}$ and $G_F=1.16637\times 10^{-5}
\text{GeV}^{-2}$ as electroweak input parameters and use Standard
Model tree level relations to compute the weak mixing angle and the
electromagnetic coupling constant.

The factorization scale is set to $\mu_F=(p_{T}^{j1} p_{T}^{j2}
p_{T}^{j3})^{1/3}$ and the renormalization to
\begin{equation}
  \alpha_s^{5}(\mu_R) = \alpha_s(p_T^{j_1}) \alpha_s(p_T^{j_2})
  \alpha_s(p_T^{j_3}) \alpha_s(p_{\Phi})^2 \, .
\end{equation}
Here, $p_T^{j_i}$ with $i=1,2,3$ denotes jets with decreasing
transverse momenta.

In the following, if not stated otherwise, we simulate effects of a
general $\mathcal{CP}$-violating Higgs boson, $\Phi$, using a toy
model. 
In general, the $Agg$ coupling is enhanced in comparison to the $Hgg$
one by factor 3/2 due to loop effects (see Lagrangian of
Eq.~\ref{eq:ggS}).  Although, in our MC program, it is possible to
modify arbitrarily the strength of Yukawa couplings being important
for general studies, in the considered model, we impose an equal
coupling strength to fermions for both, the $\mathcal{CP}$-even and
the $\mathcal{CP}$-odd parts of $\Phi$. To achieve this, it is
necessary to reduce the strength of the $\mathcal{CP}$-odd coupling by
a factor of 2/3 with respect to Eq.~(\ref{c:tb}):
\begin{align}
y_d=\frac{3}{2} \tilde{y}_d = - \frac{m_d}{v} \tan \beta \quad
\text{and} \quad y_u=\frac{3}{2} \tilde{y}_u = - \frac{m_u}{v} \cot
\beta.
\label{eq:model}
\end{align}
This set up, as we will see below, will produce a known behavior in
the differential distributions sensitive to $\mathcal{CP}$-Higgs
measurements.
The left panel of Figure~\ref{fig:massscanA} shows for different
values of $\tan \beta$ the total cross section of a pure
$\mathcal{CP}$-odd Higgs boson as a function of its mass. One can
observe that amplitudes containing both, top and bottom loop
corrections, denoted by ``t+b'' in the following, are
indistinguishable from the pure top loop contributions for $\tan
\beta=1$. Both of them demonstrate
visibly the expected threshold enhancement at a Higgs mass value
corresponding to twice of the top mass. 
In the case of bottom-quark loop dominated processes, the characteristical peak
appears well below the shown Higgs mass range. We also show results
for the effective theory approximation with and without applying
corrections to the couplings by an additional form factor (FF),
obtained from
Eq.~(2.26) of Ref.~\cite{Djouadi:2005gj}.

Within a $10\%$ deviation with respect to the full theory, the
effective theory gives accurate predictions up to Higgs masses of
$150$ GeV.

With the help of the form factor FF, similarly applied in the purely
$\mathcal{CP}$-even Higgs boson case of
Ref.~\cite{Campanario:2013mga}, the validity range is extended up to
Higgs masses of about $300 \GeV$ within a $ 10\%$
deviation. Additionally, it introduces back the threshold behavior at
$m_A=2m_t$. Beyond that validity bound, the total cross section is
overestimated up to 20$\%$ at $m_H=370$ GeV, and converges afterwards
slowly to the full theory result for the shown Higgs mass
range. Although, the form factor predicts the normalization of the
cross section for $\tan \beta = 1$ relatively well, large deviations
can be still observed in differential distributions, see
Ref.~\cite{Campanario:2013mga}.
\begin{center}
\begin{figure}[!thb]
\includegraphics[width=0.957\columnwidth]{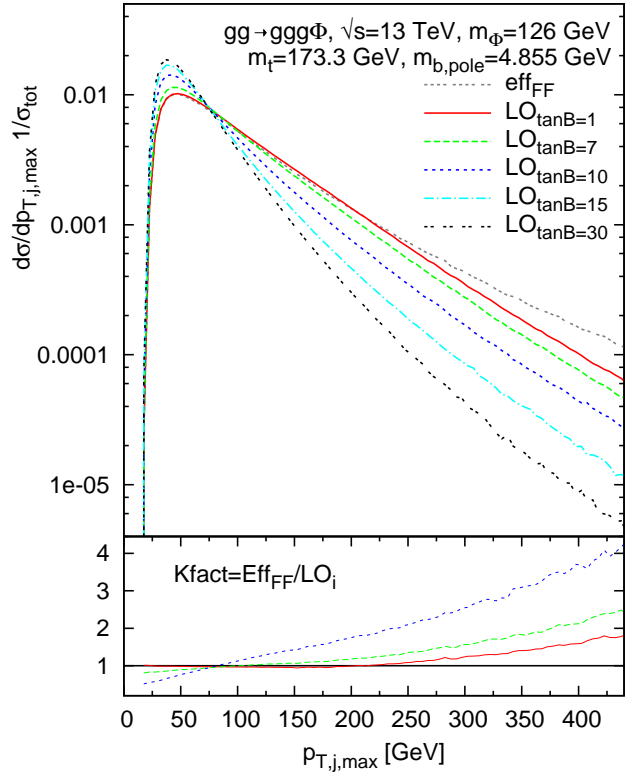}
\hfill
\caption{Transverse-momentum distributions of the harder jet generated
  within our toy-model scenario including top and bottom loop-induced
  amplitudes for different values of $\tan \beta$ and for the
  effective theory with form factors (eff$_{FF}$). The lower panel
  shows the ratios of the effective Lagrangian approach vs. the full
  theory for various $\tan \beta$ values. The inclusive cuts (IC) of
  Eq.~\eqref{ICuts} are applied.}
\label{ptjet:diff}
\end{figure}
\end{center}
\begin{center}
\begin{figure*}[!thb]
\includegraphics[width=0.902\columnwidth]{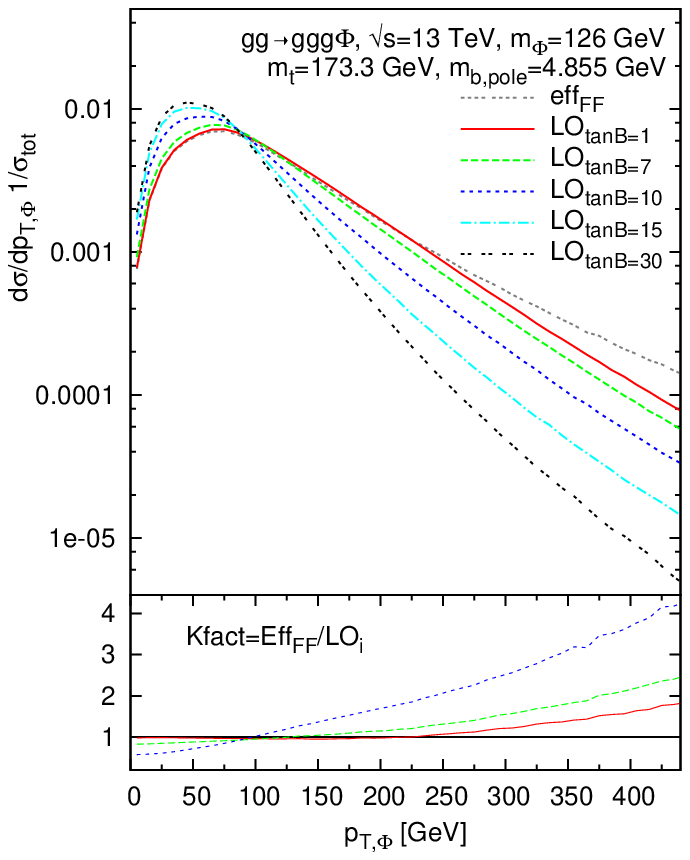}
\hfill
\includegraphics[width=0.902\columnwidth]{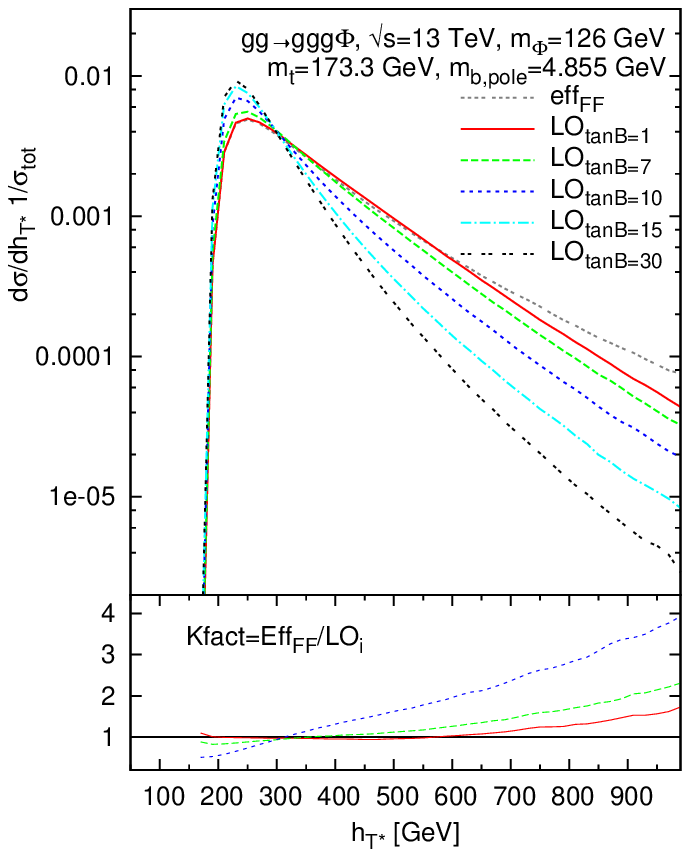}
\caption{The transverse-momentum distributions of the Higgs boson
  $\Phi$ (left panel) and the transverse scalar sum (right panel), are
  plotted. Details are described in Fig.~\ref{ptjet:diff} and in the
  text.}
\label{pthjet:diff}
\end{figure*}
\end{center}

In the right panel of Figure~\ref{fig:massscanA}, we show the total
cross section of the production of $\Phi jjj$ as a function of the
parameter $\tan \beta$ computed within our toy-model scenario for
different Higgs mass values. Similarly to $\Phi jj$ production
process~\cite{Campanario:2010mi}, the minimal cross section for small
Higgs masses is obtained near $\tan \beta \approx 7$, when $y_t
\approx y_b$ (see Eq.~(\ref{c:tb})) and both Yukawa couplings are
suppressed simultaneously in comparison to $y^{SM}$. The shift of the
minimum of $\sigma$ to larger $\tan \beta$ values with increasing
$m_{\Phi}$ can be understood in the following way: For large values of
$\tan \beta$, e.g. $\tan \beta= 30$, illustrated in the left panel of
Fig.~\ref{fig:massscanA}, the bottom-loop contributions dominate over
the suppressed top-quark contributions. However, the total cross
section decreases rapidly with rising $m_{\Phi}$ values since the
scale in the loops is now set by the heavy Higgs mass instead of the
(relatively) lighter quark mass. The suppression of bottom-loops at
large $m_{\Phi}$ implies the equality of the top- and bottom-quark
contributions, and therefore as a consequence it leads to a shifted
minimum of the distribution towards larger values of $\tan \beta$.

In the following, for a set of different $\tan \beta$ values, we
simulate effects of a general $\mathcal{CP}$-violating Higgs sector at
the LHC using the previously described toy model scenario and show
differential distributions for several phenomenologically interesting
observables for $\Phi jjj$ production with a Higgs mass fixed at $126
\GeV$.
Fig.~\ref{ptjet:diff} shows the differential distribution of the
hardest jet. For large values of $\tan \beta$, bottom loop corrections
dominate, and hence, provide a strong impact on the spectrum. For
$p_{T,j}>m_b$, the large scale of the kinematic invariants leads to an
additional suppression of bottom-loop induced sub-amplitudes compared
to the heavy quark effective theory. It is e.g clearly visible for
$\tan \beta=10$ at $400 \GeV$, where Kfact (lower panel) shows a 4-times
overestimated prediction of the heavy theory approach.
For $\tan \beta=1$, the effective theory approximation describes
efficiently the full theory prediction within $10\%$ accuracy up to
$p_{T}^{j\text{max}} < 200$ GeV. Beyond that regime, differences start
to increase and deviations up to $100\%$ are found.
Hence, these facts stress the limited predictive power of
the heavy-top quark limit approximation in scenarios beyond the SM.
Similar properties are found in Fig.~\ref{pthjet:diff}, where the left
panel illustrates the differential distribution of the transverse Higgs boson
momentum. The right panel shows the transverse scalar sum of the system,
oftenly used in the framework of new physics searches,  defined as  $H_T= \sum _i p_T^{j_i} + \sqrt{p_{T,\Phi}^2+
  M_{\Phi}^2}$.
\begin{center}
\begin{figure*}[!thb]
\includegraphics[width=0.903\columnwidth]{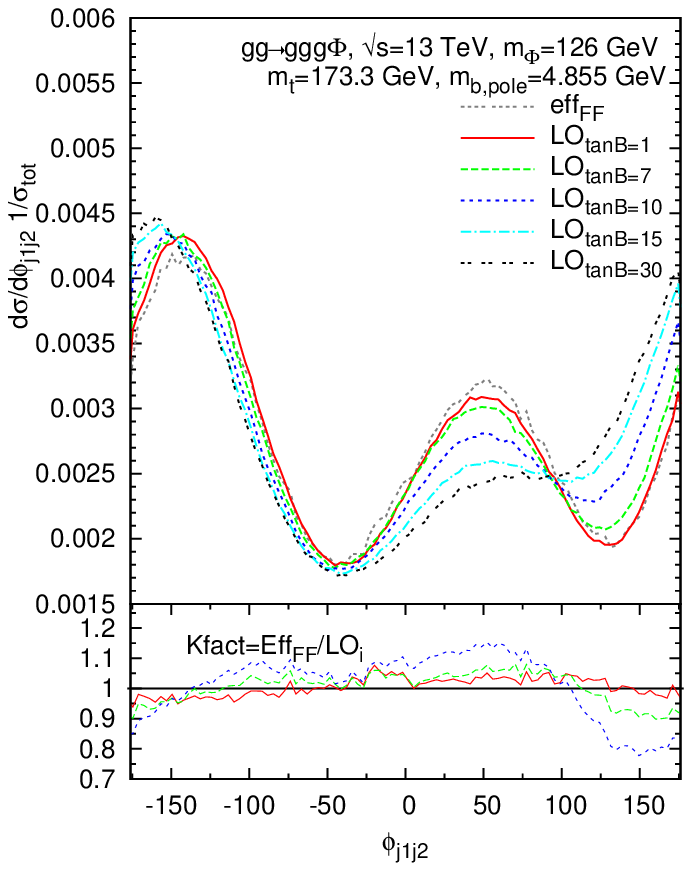}
\hfill
\includegraphics[width=0.903\columnwidth]{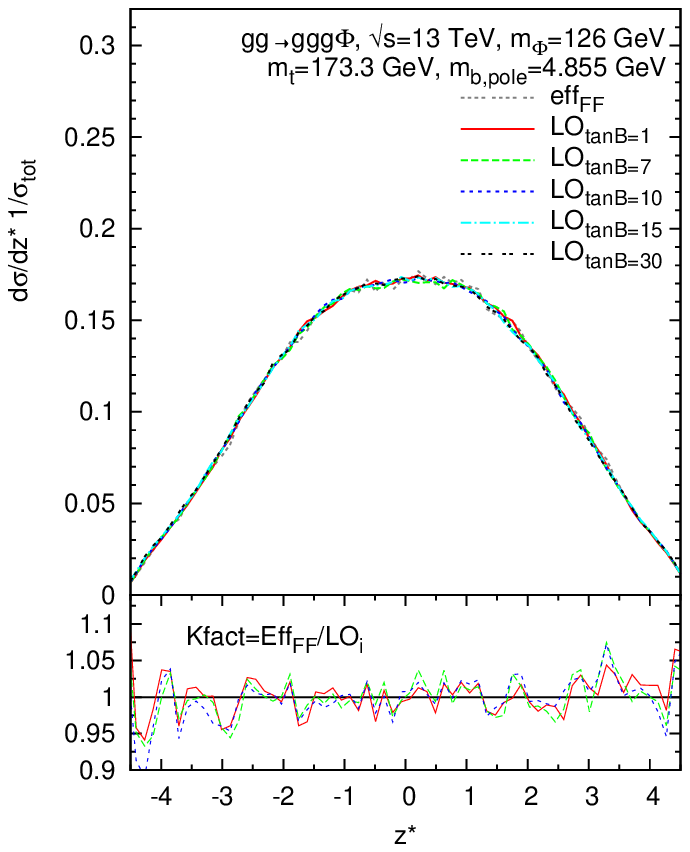}
\caption{Left: azimuthal angle correlation $\phi_{j1j2}$ of the two
  harder jets with applied ICphi cuts of Eq.~(\ref{ICphi}). Right:
  $z^*$, the normalized centralized rapidity distribution of the third
  jet w.r.t. the tagging jets using the VBF cuts of
  Eq.~(\ref{eq:vbfcuts}). Further, details are described in
  Fig.~\ref{ptjet:diff} and in the text.}
\label{phi:diff}
\end{figure*}
\end{center}
The azimuthal angle distribution is sensitive to the
$\mathcal{CP}$-character of the Higgs coupling to fermions. In
Ref.~\cite{Campanario:2010mi}, it was proven for $\Phi jj$
production that the softening effects observed in the transverse
momentum distributions due to bottom-loop corrections did not modify
the jet azimuthal angle correlations predicted by the effective theory
approximation. In this letter, the presence of the additional third
jet rises the question whether soft radiation can distort these
predictions. We follow the definition of Ref.~\cite{Hankele:2006ma} of
the azimuthal angle distribution between the more forward and the more
backward of the two tagging jets. To increase the sensitivity to the
$\mathcal{CP}$-structure of the Higgs couplings, a modification of the
inclusive set of cuts is applied,
\begin{align} 
\label{ICphi} p_{T}^{j_i} > 30 \ \text{GeV}\;, \hspace{0.1cm} |\eta_j| < 4.5\;,
\hspace{0.1cm} R_{jj} > 0.6\;, \hspace{0.1cm} \Delta \eta_{jj}>3 \;.
\end{align}
We refer to them as ICphi set of cuts in the following. The effective
theory approach showed a phase shift of the $\phi_{jj}$ distribution
by an angle $\alpha$ which is given by the relative strength of the
$\mathcal{CP}$-even and $\mathcal{CP}$-odd couplings. Taking into
account the relative enhancement of the pure $\mathcal{CP}$-odd
coupling due to loop effects (see Eq.~(\ref{eq:ggS})), the phase shift
angle is given by~\cite{Hankele:2006ma},
\begin{align}
\tan \alpha = \frac{3}{2} \frac{\tilde{y}_q}{y_q} \,.
\end{align}
In our toy model scenario, Eq.~(\ref{eq:model}), we assume $y_q=3/2
\tilde{y}_q$, and hence, the minima are shifted to $\alpha=
45^o,135^o$ degrees. This can be seen in Figure~\ref{phi:diff}, where
we simulate effects of a general $\mathcal{CP}$-violating Higgs sector
at the LHC for a set of different $\tan \beta$ values, illustrated
with the help of the normalized $\phi_{jj}$-distributions. The
effective theory approximation reproduces accurately the shape of the
$\phi_{jj}$ distribution. Whereas in the full theory, the azimuthal
angle distributions receives kinematic distortions which are caused by
kinematical effects due to both, the balance of the transverse
momentum of the jets and the Higgs boson, and the softer momentum
spectrum of the jets and the Higgs boson for high values of $\tan
\beta$ (Fig.\ref {ptjet:diff} and \ref{pthjet:diff}) where bottom loop
contributions dominate.
Using typical vector fusion cuts,
\begin{equation}
m_{j1j2} > 600 \GeV,~~ |y_{j1}-y_{j2}|> 4,~~  y_{j1} \cdot y_{j2} < 0 \;,
\label{eq:vbfcuts}
\end{equation}
 we show the normalized centralized rapidity distribution of the third
 jet with respect to the tagging jets, $z^* = (y_3 - 1/2 (y_1 + y_2
 ))/|y_1 - y_2|$. This variable reflects the nature of VBF processes
 involving the fusion of electro-weak Gauge bosons. In EW Hjjj
 production~\cite{Campanario:2013fsa}, one can clearly observe how the
 third jet tends to accompany one of the leading jets appearing at 1/2
 and -1/2 respectively. Additionally, due to its color singlet nature,
 there is almost no jet activity in the rapidity gap region (minimum
 at $z^*=0$) between the two leading tagging jets. In our case, it
 shows the typical behavior of a QCD induced process, and the rapidity
 gap between the two jets is filled up by at least a third jet due to
 additional gluon radiation. Furthermore, the $z^*$-distribution is
 insensitive to the change of the Higgs couplings to fermions by the
 model parameter $\tan \beta$.
\section{Summary}
\label{sec:summ}
In this letter, we have presented first results for the gluon fusion
loop-induced sub-process $gg \to ggg \Phi$ at the LHC, where $\Phi$
corresponds to a general $\mathcal{CP}$-violating Higgs
boson. Interference effects between loops with top- and bottom-quarks
as well as between $\mathcal{CP}$-even and $\mathcal{CP}$-odd
couplings of the heavy quarks were fully taken into account.

The stability of the numerical results is guaranteed by a suitable
application of Ward identities and quadruple precision, which are 
adecuate even for bottom dominated configurations.

Using a toy model scenario, we have presented effects of bottom-quark
loop contributions which can lead to visible distortions in the
differential distributions of important observables for large values
of $\tan \beta$. 

Operating at a center of mass energy of $\sqrt{s}=13$ TeV, for small
values of $\tan \beta$, up to Higgs masses of 290 GeV and for small
transverse momenta $p_{T}^{j\text{max}} \lesssim 290$ GeV, the
effective Lagrangian approximation including the form factor
correction gives accurate results and can be used as a numerically
fast alternative for phenomenological studies.
No restriction was found in the validity of the invariant mass of the
dijet system of the leading jets (not shown) for small values of $\tan
\beta$.
The shape of the azimuthal angle distribution is well described
by the effective theory. However, distortions in the shape appear
for increasing values of $\tan \beta$.
A detailed description of the full process will be given in a
forthcoming publication.
This process will be made publicly available as part of the VBFNLO
program.

\section*{Acknowledgments} 
\noindent It is a pleasure to thank Dieter Zeppenfeld for fruitful
discussions during the development of this project.
We acknowledge the support from the Deutsche Forschungsgemeinschaft
via the Sonderforschungsbereich/Transregio SFB/TR-9 Computational
Particle Physics. FC is funded by a Marie Curie fellowship
(PIEF-GA-2011-298960) and partially by MINECO (FPA2011-23596) and by
LHCPhenonet (PITN-GA-2010-264564).
MK acknowledges support by the Grid Cluster of the RWTH-Aachen.
\appendix

\bibliographystyle{h-physrev}
\bibliography{ggA}

\begin{thebibliography}{10}

\bibitem{Plehn:2001nj}
T.~Plehn, D.~L. Rainwater, and D.~Zeppenfeld,
\newblock Phys.Rev.Lett. {\bf 88}, 051801 (2002), hep-ph/0105325.

\bibitem{Djouadi:2005gi}
A.~Djouadi,
\newblock Phys.Rept. {\bf 457}, 1 (2008), hep-ph/0503172.

\bibitem{Djouadi:2005gj}
A.~Djouadi,
\newblock Phys.Rept. {\bf 459}, 1 (2008), hep-ph/0503173.

\bibitem{Cox:2010ug}
B.~E. Cox, J.~R. Forshaw, and A.~D. Pilkington,
\newblock Phys.Lett. {\bf B696}, 87 (2011), 1006.0986.

\bibitem{Coleppa:2012eh}
B.~Coleppa, K.~Kumar, and H.~E. Logan,
\newblock Phys.Rev. {\bf D86}, 075022 (2012), 1208.2692.

\bibitem{Freitas:2012kw}
A.~Freitas and P.~Schwaller,
\newblock Phys.Rev. {\bf D87}, 055014 (2013), 1211.1980.

\bibitem{Englert:2012ct}
C.~Englert, M.~Spannowsky, and M.~Takeuchi,
\newblock JHEP {\bf 1206}, 108 (2012), 1203.5788.

\bibitem{Harlander:2013oja}
R.~V. Harlander and T.~Neumann,
\newblock Phys.Rev. {\bf D88}, 074015 (2013), 1308.2225.

\bibitem{Chang:2013cia}
W.-F. Chang, W.-P. Pan, and F.~Xu,
\newblock Phys.Rev. {\bf D88}, 033004 (2013), 1303.7035.

\bibitem{Djouadi:2013qya}
A.~Djouadi and G.~Moreau,
\newblock (2013), 1303.6591.

\bibitem{Aad:2012tfa}
ATLAS Collaboration, G.~Aad {\em et~al.},
\newblock Phys.Lett. {\bf B716}, 1 (2012), 1207.7214.

\bibitem{Chatrchyan:1471016}
S.~Chatrchyan {\em et~al.},
\newblock Phys. Lett. B {\bf 716}, 30 (2012).

\bibitem{DelDuca:2001eu}
V.~Del~Duca, W.~Kilgore, C.~Oleari, C.~Schmidt, and D.~Zeppenfeld,
\newblock Phys.Rev.Lett. {\bf 87}, 122001 (2001), hep-ph/0105129.

\bibitem{Odagiri:2002nd}
K.~Odagiri,
\newblock JHEP {\bf 0303}, 009 (2003), hep-ph/0212215.

\bibitem{Hankele:2006ma}
V.~Hankele, G.~Klamke, D.~Zeppenfeld, and T.~Figy,
\newblock Phys.Rev. {\bf D74}, 095001 (2006), hep-ph/0609075.

\bibitem{Klamke:2007cu}
G.~Klamke and D.~Zeppenfeld,
\newblock JHEP {\bf 0704}, 052 (2007).

\bibitem{Hagiwara:2009wt}
K.~Hagiwara, Q.~Li, and K.~Mawatari,
\newblock JHEP {\bf 0907}, 101 (2009), 0905.4314.

\bibitem{Campanario:2010mi}
F.~Campanario, M.~Kubocz, and D.~Zeppenfeld,
\newblock Phys.Rev. {\bf D84}, 095025 (2011), 1011.3819.

\bibitem{DelDuca:2006hk}
V.~Del~Duca {\em et~al.},
\newblock JHEP {\bf 0610}, 016 (2006).

\bibitem{DelDuca:2008zz}
V.~Del~Duca,
\newblock Acta Phys.Polon. {\bf B39}, 1549 (2008).

\bibitem{Andersen:2010zx}
J.~R. Andersen, K.~Arnold, and D.~Zeppenfeld,
\newblock JHEP {\bf 1006}, 091 (2010), 1001.3822.

\bibitem{Campbell:2006xx}
J.~M. Campbell, R.~K. Ellis, and G.~Zanderighi,
\newblock JHEP {\bf 0610}, 028 (2006), hep-ph/0608194.

\bibitem{vanDeurzen:2013rv}
H.~van Deurzen {\em et~al.},
\newblock Phys.Lett. {\bf B721}, 74 (2013), 1301.0493.

\bibitem{Campanario:2013mga}
F.~Campanario and M.~Kubocz,
\newblock Phys.Rev. {\bf D88}, 054021 (2013), 1306.1830.

\bibitem{Cullen:2013saa}
G.~Cullen {\em et~al.},
\newblock Phys.Rev.Lett. {\bf 111}, 131801 (2013), 1307.4737.

\bibitem{Arnold:2008rz}
K.~Arnold {\em et~al.},
\newblock Comput.Phys.Commun. {\bf 180}, 1661 (2009), 0811.4559.

\bibitem{Arnold:2011wj}
K.~Arnold {\em et~al.},
\newblock (2011), 1107.4038.

\bibitem{Arnold:2012xn}
K.~Arnold {\em et~al.},
\newblock (2012), 1207.4975.

\bibitem{Hagiwara:1985yu}
K.~Hagiwara and D.~Zeppenfeld,
\newblock Nucl.Phys. {\bf B274}, 1 (1986).

\bibitem{Hagiwara:1988pp}
K.~Hagiwara and D.~Zeppenfeld,
\newblock Nucl.Phys. {\bf B313}, 560 (1989).

\bibitem{Campanario:2011cs}
F.~Campanario,
\newblock JHEP {\bf 1110}, 070 (2011), 1105.0920.

\bibitem{Passarino:1978jh}
G.~Passarino and M.~Veltman,
\newblock Nucl.Phys. {\bf B160}, 151 (1979).

\bibitem{Denner:2005nn}
A.~Denner and S.~Dittmaier,
\newblock Nucl.Phys. {\bf B734}, 62 (2006), hep-ph/0509141.

\bibitem{'tHooft:1978xw}
G.~'t~Hooft and M.~Veltman,
\newblock Nucl.Phys. {\bf B153}, 365 (1979).

\bibitem{Denner:1991qq}
A.~Denner, U.~Nierste, and R.~Scharf,
\newblock Nucl.Phys. {\bf B367}, 637 (1991).

\bibitem{Alwall:2007st}
J.~Alwall {\em et~al.},
\newblock JHEP {\bf 0709}, 028 (2007), 0706.2334.

\bibitem{Alwall:2011uj}
J.~Alwall, M.~Herquet, F.~Maltoni, O.~Mattelaer, and T.~Stelzer,
\newblock JHEP {\bf 1106}, 128 (2011), 1106.0522.

\bibitem{Campanario:2012bh}
F.~Campanario, Q.~Li, M.~Rauch, and M.~Spira,
\newblock (2012), 1211.5429.

\bibitem{Pumplin:2002vw}
J.~Pumplin {\em et~al.},
\newblock JHEP {\bf 0207}, 012 (2002).

\bibitem{Spira:1997dg}
M.~Spira,
\newblock Fortsch.Phys. {\bf 46}, 203 (1998), hep-ph/9705337.

\bibitem{Vermaseren:1997fq}
J.~Vermaseren, S.~Larin, and T.~van Ritbergen,
\newblock Phys.Lett. {\bf B405}, 327 (1997), hep-ph/9703284.

\bibitem{Campanario:2013fsa}
F.~Campanario, T.~Figy, S.~Platzer, and M.~Sjodahl,
\newblock Phys.Rev.Lett. {\bf 111}, 211802 (2013), 1308.2932.

\end{thebibliography}

\end{document}